\documentclass[12pt]{article}

\usepackage{amsfonts,amssymb,epsf}

\evensidemargin=\oddsidemargin
\newcommand{\dimaplota}{\raisebox{0pt}{\dima M}}
\newcommand{\dimaplotb}{\rectangle <1\rc> <1\rc> }
\newcommand{\dimaplotc}{\raisebox{-1pt}{\dima O}}
\newcommand{\dimaplotd}{\rule{0.3em}{0.3em}}
\def\rectangle <#1> <#2> {\setbox0=\hbox{}\wd0=#1\ht0=#2\frame{\box0}}
\font\dima=msam7
\newlength{\rc}
\newlength{\FigWidth} 
\newlength{\FigHeight} 
\makeatletter
\def\eqalign#1{\null\vcenter{\def\\{\cr}\openup\jot\m@th%
  \ialign{\strut$\displaystyle{##}$\hfil&$\displaystyle{{}##}$\hfil%
      \crcr#1\crcr}}\,}%
\def\@cite#1#2{{#1\if@tempswa , #2\fi}}
\def\@biblabel#1{}
\makeatother

\begin{document}

\title{Quasars in a merger model: comparison with the
observed luminosity function}
\author{D.~S.~Krivitsky, V.~M.~Kontorovich\\
\it Institute of Radio Astronomy, Kharkov}
\date{January 17, 1998}
\maketitle

\begin{abstract}
Connection between the quasar luminosity function and galaxy mass
function is investigated in the framework of a phenomenological
approach which relates AGN formation to galaxy mergers. Quasars
are assumed to be short-lived, the luminosity of a quasar is
controlled by the masses and angular momenta of the merged
galaxies which have formed the quasar, and the amount of gas in
them (the masses and momenta determine the quantity of mass which
loses its angular momentum and can fall to the center). The
proposed model can explain the shape and evolution of the quasar
luminosity function, and allows us to estimate the parameters: the
fraction of matter which falls into the center $\eta$ (which seems
to be related to the quantity of gas in the galaxies) and
$\kappa$ (an average density contrast in the regions
where quasars form). The obtained values of $\kappa$
vary from $\sim4\mbox{--}7$ at $z=0.5$ to $\sim1\mbox{--}2$ at
$z=2$, $\eta$ vary from a few per cent at $z=0.5$ to a few tens
per cent or even values close to 1 at $z=2$. In contrast to the
cases considered earlier by the authors, the Eddington limit
which, probably, can be exceeded in quasars plays an essential
role.
\end{abstract}

\section{Introduction}
In present time, it has been established rather reliably that
galaxy interaction (in particular, merging) correlates with
the activity of galactic nuclei, at least, for powerful objects
(see, e.g., reviews by Heckman, \cite{hek}; Kontorovich,
\cite{kont94}, and references therein). There is still some
uncertainty for less powerful AGN, such as Seyfert galaxies.
However, in the spirit of the unified AGN scheme (Antonucci,
\cite{anton}) we may suppose that here we also deal with
interaction.

Note, that such bright objects as quasars are a sort of markers:
the quasar formation epoch which is often identified with the
well-known cutoff in their distribution at $z=z_{\rm cr} \approx
2.5$ (Schmidt et~al., \cite{schmidt}; Shaver, \cite{shaver}) may be
also the epoch of massive galaxy formation due to mergers of less
massive blocks (see Kats et~al., \cite{kkk92}; Kontorovich et~al.,
\cite{jetplet})\footnote{The more so, as the merger process may be
``explosive'', see Cavaliere et~al. (\cite{ccm91}); Kontorovich
et~al. (\cite{jetplet}).} Note, that reality of the cutoff in the
$z$-dependence of quasar density is confirmed by the counts of
radio sources (see the review by Peakock, \cite{pea} and the paper
by Artyukh and Tyulbashev, \cite{at}).

Last data from the Hubble Space Telescope seem to confirm this
point of view. Observations of galaxy formation of blocks (merging
process) in the redshift range $2.6<z<3.9$ allows Clements and
Couch (\cite{CC}) to conclude that, possibly, an epoch of
({\it massive}, K.~K.) galaxy formation has been discovered.
Observations of subgalactic blocks at $z=2.39$ (Pascarelle et~al.,
\cite{PAS}) and their relation to galaxy formation were discussed
in details for the rich group which was discovered in connection
with a faint blue galaxies investigation program (see also other
works of the same group:  Windhorst et~al., \cite{vind93};
\cite{vind95}, and references therein). Recent source counts also
allow to explain their rise to the past (for the standard
cosmology and critical density) by evolution of the number of
galaxies due to their mergers, assuming that this epoch
corresponds to $z \approx 2$ (Metcalfe et~al., \cite{count}; see
also Bender and Davis, \cite{GalEvol}).

On the other hand, direct observations of close quasar host
galaxies by the HST (Bahcall et~al., \cite{BKS94}; \cite{BKS95})
gave a remarkable confirmation of the direct connection between
the activity and the galaxy interaction and merging. In
particular, in the case of PKS~2349 quasar host galaxy, a LMC-size
sinking satellite galaxy was discovered\footnote{In some sense, it
is rather the absolutely unperturbed elliptical host galaxies,
discovered by the same group, to be a puzzle.}

So, there are strong reasons to continue investigation of the
relation between the activity and galaxy mergers (Kontorovich,
Krivitsky, \cite{kk95pazh}) and perform a more detailed comparison
between the observations and the phenomenological scheme which was
proposed earlier and improved below. We assume that pairwaise
mergers of galaxies, taking place due to their
gravitational interaction and tidal forces, are the factor
which triggers
activity. In this approach, quasar luminosity function (LF) and
galaxy mass function (MF) turn out to be related. Below we shall
assume that galaxy MF is known and, thus, shall not make any
assumptions about the mechanism of its formation. The activity, in
turn, is controlled by mergers.

In this work we shall found the quasar LF and the values of
parameters (the density contrast and the parameter which
determines probably the amount of gas in galaxies) for which it
agrees with observational data in the redshift range $0.5 \lesssim
z \lesssim 2$; we shall analyze also the connection between galaxy
MF and quasar LF asymptotical behavior.

\section{Discussion of the model}
\label{discus}
The probable cause of the correlation between the galaxy interaction
(in particular, merging) and their nuclear activity is that the
interaction leads to redistribution of the angular momentum and,
therefore, some part of matter (probably, gas) gets
into the central region and gives material for accretion (see,
e.g., Hernquist, Barnes, \cite{gas-to-center}).

In the proposed earlier model (Kats, Kontorovich, \cite{kk91pazh};
Kontorovich, Krivitsky, \cite{kk95pazh}) which describes appearing
of activity due to mergers, the falling of matter to the center was
assumed to be related to compensation of a part of the angular
momentum at the merger. Below this assumption will be considered
as a special case. According to this approach, the most important
parameters of the problem are galaxy masses ${\frak M}$,
angular momenta ${\bf S}$, and the amount of gas.

We shall assume that the luminosity of an active galaxy formed as
a result of merging between two galaxies is controlled by their
masses ${\frak M}_{1,2}$ and momenta ${\bf S}_{1,2}$, as well as
the collision orbital momentum~${\bf J}$: $L=L({\frak M}_1,{\frak
M}_2,{\bf S}_1,{\bf S}_2,{\bf J})$ (taking into account the amount
of gas will be discussed below). To compute this function, a
detailed theory is needed, which would deal with a very
complicated multi-step process due to which a part of matter lose
its momentum and gets into the center after the merger. However,
we shall restrict ourselves by a simplified phenomenological
approach. The number of active galaxies formed per unit time and
unit luminosity range is, obviously, expressed as
\begin{equation}
\eqalign
{I(L)=\int f({\frak M}_1,{\bf S}_1)f({\frak M}_2,{\bf
S}_2)U({\frak M}_1,{\frak M}_2) F({\bf J})
\times\\\qquad
\delta(L-L({\frak M}_1,{\frak M}_2,{\bf S}_1,{\bf S}_2,{\bf
J}))\,d{\frak M}_1\,d{\frak M}_2 \,d^3S_1\,d^3S_2\,d^3J,}
\label{ilisotr}
\end{equation}
Here $f({\frak M},{\bf S})$ is the galaxy mass and angular
momentum distribution function; $U({\frak M}_1,{\frak M}_2)$ is a
characteristic of the probability of a merger between galaxies
with masses ${\frak M}_1$ and ${\frak M}_2$ (in general, $U$
depends not only on masses, but also on momenta, however, this
dependence seems to be less essential and will not be taken into
account in this work); $F({\bf J})$ is the angular momentum
distribution function. Thus, given $f({\frak M},{\bf S})$, we can
found the rate of active objects formation, as a function of
their luminosity~$I(L)$.
%$

Next, it is possible to relate $I(L,t)$ to the active nuclei LF
$\phi(L,t)$. To do it, we have to make some assumptions about the
evolution (i.e, in our case, the light curves) of active nuclei
forming by mergers. Thus, if we assume that the light curve is
step-shaped, with the average duration $t_{\rm act}$, then
$\phi(L,t)$ can be described by the model equation
${\textstyle\partial\phi\over\textstyle\partial t}=I-\phi/t_{\rm
act}$ (here $t_{\rm act}$ may, in general, depend on~$L$). Another
possible case: if we assume that the luminosity of an active
galaxy decreases exponentially, with the e-fold time $t_{\rm
act}$, then $\phi(L,t)$ is described by the equation
${\textstyle\partial\phi\over\textstyle\partial
t}-{\textstyle\partial\over\textstyle\partial L}\left({\textstyle
L\phi\over\textstyle t_{\rm act}}\right)=I$ (an analogue of
the continuity equation in the luminosity space; cf. Cavaliere
et~al., \cite{cgv}). In the latter case, large lifetime ($t_{\rm
act}\gtrsim10^9$ years) and small AGN formation rate
($I\ll\left|{\textstyle\partial\over\textstyle\partial L}\left(
{\textstyle L\phi\over\textstyle t_{\rm act}}\right)\right|$)
corresponds to the ``luminosity evolution'': changing $\phi$
reflects, mainly, reducing luminosity of the existing objects. The
case of small lifetime ($t_{\rm act}\lesssim10^8$ years) for both
equations corresponds to existence of many AGN generations, which
change each other in the course of the Universe evolution
(``number evolution''). In this paper we shall consider the case
of number evolution (see more detailed discussion below, in
section~\ref{sec3}). Assuming $t_{\rm act}$ much less than a
characteristic time of $I(L,t)$ changing, we have for the former
equation\footnote{The solutions of both equations can be easily
written in an explicit form for arbitrary $t_{\rm act}$, but below
only the case of small $t_{\rm act}$ will be of interest for us.
It is possible also to write an explicit solution for arbitrary
(not necessarily exponential) light curve.}
\begin{equation}
\phi(L,t)\approx t_{\rm act}I(L,t);
\label{ur1}
\end{equation}
and, for the latter equation
\begin{equation}
\phi(L,t)\approx\frac{\textstyle t_{\rm act}}{\textstyle
L}\int_L^\infty I(L',t)\,dL'.
\label{ur2}
\end{equation}
Qualitatively these two expressions are very similar: if $I(L)$ has
a power law region and an exponential decrease region, then the
shape of $\phi(L)$ is approximately the same as the one of $I(L)$
(we shall not consider
the case when $t_{\rm act}$ depends\footnote{In particular,
the Eddington time does not depend of galaxy parameters.}
on~$L$, though it can be easily done if there appear some
observational data about a dependence of $t_{\rm act}$ on $L$ or
some other galaxy parameters).
So, in fact, we shall investigate the source $I(L)$ in
the equation for the LF.

In our previous works we considered an expression for
$f({\frak M},{\bf S})$, which corresponds to the ``anisotropic''
momentum distribution. This distribution appeared in Kats,
Kontorovich (\cite{kk90jetp}) as a solution of the generalized
Smoluchowski kinetic equation (which describes galaxy mergers)
without allowance for the orbital momentum, if there is some initial
anisotropy. In this case, the initial anisotropy is amplified in
the course of time, the momentum distribution tends to a
$\delta$-function, and the momentum of a galaxy is proportional to
its mass:
\begin{equation}
f({\frak M},{\bf S})\approx\Phi({\frak M})\delta\left
({\bf S}-\frac{{\bf S}_0{\frak M}}{{\frak M}_0}\right),
\label{anis}
\end{equation}
where $\Phi({\frak M})$ is the MF. Distribution (\ref{anis}) is
useful from the methodical point of view, because the asymptotical
behavior of $I(L)$ can be computed analytically for it. However,
from the astrophysical point of view, isotropic distribution
\begin{equation}
f({\frak M},{\bf
S})=\Phi({\frak M})\frac1{\left(\frac{2\pi}3\overline{S^2({\frak M})}\right)^{3/2}}
\exp\left(-\frac32S^2/\overline{S^2({\frak M})}\right),
\label{isotr}
\end{equation}
which will be considered in this work, is more interesting.

The mass dependence of an average mass to luminosity ratio for
normal galaxies is rather weak. We shall neglect this
dependence\footnote{Evolution of this ratio which reflects
evolution of star formation determined, in particular, by
mergers may be very important, cf. Madau (\cite{madau}).} and take
$\frac {{\frak M}}L\sim10$. Then the MF just coincides with the
LF (except for the normalization). Below, in section~\ref{sec2},
for computing $I(L)$, we shall take the MF in Schechter's form
\begin{equation}
\Phi({\frak M})=\Phi_0{\frak M}^{-\alpha}e^{-{\frak M}/\mu}.
\label{sheh}
\end{equation}
The index $\alpha\approx1$ for field galaxies and
$\alpha\gtrsim1.25$ for clusters (see, e.g., Binggeli et~al.,
\cite{binggeli}; Loveday et~al., \cite{lovd1}). As the integral
which expresses the total number of galaxies for (\ref{sheh})
diverges, we shall assume $\Phi=0$ at small masses ${\frak
M}<{\frak M}_0$. Observational data obtained in recent years testify to
possible steepening of $\Phi$ at small masses. This steepening
will be taken into account in section \ref{sec3}.

We will use a rather common merger criterion: (i)~minimal distance
between the colliding galaxies is less that the sum of their radii
$(R_1+R_2)$; (ii)~the relative velocity at infinity $v$ is less
than $v_{\rm g}=\sqrt{\frac{2G({\frak M}_1+{\frak
M}_2)}{R_1+R_2}}$.  Then the merger cross-section (taking into
account gravitational focusing) is
\label{sigma}$\sigma=\pi(R_1+R_2)^2(1+v_{\rm g}^2/v^2)$
for $v<v_{\rm g}$ and
$\sigma=0$ for $v>v_{\rm g}$. It results in the following
expression for~$U$:
\begin{equation}
U({\frak M}_1,{\frak M}_2)=\overline{\sigma
v}\approx\cases{c_2({\frak M}_1+{\frak M}_2)^2,&${\frak M}\ll
{\frak M}_{\rm b}$\cr c_{1+\beta}({\frak M}_1+{\frak
M}_2)\*({\frak M}_1^\beta+{\frak M}_2^\beta),&${\frak M}\gg {\frak
M}_{\rm b}$}
\label{U}
\end{equation}
(cf. Kats, Kontorovich, \cite{kk90jetp}; Krivitsky, Kontorovich,
\cite{kk97}; Cavaliere, Menci, \cite{cmlanl}). Here
the bar means an average over
velocities; galaxy radius $R$ relates to the mass as
$R=C{\frak M}^\beta$
($\beta=1/3$ corresponds to constant density, $\beta=1/2$ to the
observational Faber---Jackson and Tully---Fisher laws);
$c_2=(9/2)(3\pi)^{1/2}G^2/v_{\rm
rms}^3$, $c_{1+\beta}=2(3\pi)^{1/2}CG/v_{\rm rms}$,
${\frak M}_{\rm b}\sim(Cv_{\rm rms}^2/G)^{1/(1-\beta)}$.
For the function $U({\frak M}_1,{\frak M}_2)$, it is convenient to
introduce its homogeneity power $u$ and exponents $u_{1,2}$ which
describe its asymptotics for very different masses:
\begin{equation}
U\propto {\frak M}_1^{u_1}{\frak M}_2^{u_2}, \qquad {\frak M}_1\ll
{\frak M}_2,\quad u_1+u_2=u.
\label{u12}
\end{equation}
Obviously, for (\ref{U}) $u_1=0$, $u_2=u=2$ if ${\frak M}\ll
{\frak M}_{\rm b}$ and $1+\beta$ if ${\frak M}\gg {\frak M}_{\rm
b}$. Note that it is the parameters $u_{1,2}$ (that is the
asymptotical behavior of~$U$) to determine the
asymptotical behavior of $I(L)$.

\section{Asymptotics and the relation between indices}
\label{sec2}
Given the asymptotics of $L({\frak M}_1,{\frak
M}_2,{\bf S}_1,{\bf S}_2,{\bf J})$, $U({\frak M}_1,{\frak M}_2)$
and $f({\frak M},{\bf S})$, it is possible to find the
asymptotical behavior of~$I(L)$. In particular, the model
predicts that $I(L)$ has a power-law region, the slope of which
depends on the slope of the galaxy MF power-law region.

First we shall consider the simplest variant: ``anisotropic''
momentum distribution (\ref{anis}) without taking into account
the orbital momentum (${\bf J}=0$). In this case momenta can be
expressed in terms of masses (${\bf S}\propto{\frak M}$) and, so,
$L=L({\frak M}_1,{\frak M}_2)$. We shall assume that
$L({\frak M}_1,{\frak M}_2)$ has power-law asymptotical behavior
at ${\frak M}_1\ll {\frak M}_2$:
\begin{equation}
L({\frak M}_1,{\frak M}_2)\propto {\frak M}_1^{\lambda_1}{\frak
M}_2^{\lambda_2},\qquad {\frak M}_1\ll {\frak
M}_2,
\label{2.1}
\end{equation}
i.e., $L({\frak M}_1,{\frak M}_2)$ can be expressed as an
asymptotical power series with respect to both arguments. The
 right-hand part of (\ref{2.1}) is a homogeneous function of power
$\lambda=\lambda_1+\lambda_2$.  Sewing together the asymptotics
for ${\frak M}_1\ll {\frak M}_2$ and ${\frak M}_1\gg {\frak M}_2$,
we shall assume below that $L({\frak M}_1,{\frak M}_2)$ is a
homogeneous function of power~$\lambda$ in the whole range of
${\frak M}_1$, ${\frak M}_2$. Expression (\ref{2.1}) is analogous
to (\ref{u12}) for $U({\frak M}_1,{\frak M}_2)$. Knowing
$\lambda_{1,2}$ is enough to find the slope of the power-law
intermediate asymptotics of (\ref{ilisotr}).

Expression
\begin{equation}
L=B\Delta m,\quad B=\varepsilon\eta c^2t_{\rm ac}^{-1},
\quad\Delta m=m_1+m_2-m,\quad m=S/\sqrt{G{\frak M}R},
\label{1.1}
\end{equation}
which was considered in Kats, Kontorovich (\cite{kk91pazh});
Kontorovich, Krivitsky (\cite{kk95pazh}) is a particular case of
(\ref{2.1}), corresponding to $\lambda_2=0$,
$\lambda_1=\lambda>0$. Here $\Delta m$ is the mass which has lost
its momentum due to momentum compensation at the merger;
$\varepsilon$ the accretion effectivity; $\eta$ the fraction of
$\Delta m$, which gets to the central black hole; $t_{\rm ac}$
the accretion time (we shall assume $t_{\rm
ac}=t_{\rm act}$); $c$ is the light speed. Note that, though
expression (\ref{1.1}) was based on an oversimplified scheme of
the origin of activity, the assumption $\lambda_2=0$,
$\lambda_1>0$ is much more general and seems to be rather
plausible even without any connection with model (\ref{1.1}). Its
physical sense is that when a massive galaxy merges with a
low-mass one, the luminosity is determined mainly by the latter
mass. Thus, results obtained from (\ref{1.1}) are more
general than the model (\ref{1.1}). In this case, the slope of the
power-law region is determined by equations (15) and (16) from the
cited above work by Kontorovich and Krivitsky. The opposite case,
$\lambda_1=0$, $\lambda_2>0$ (i.e., the luminosity is determined,
mainly, by the more massive galaxy) was considered (equation
(18) in the same work) in connection with the situation when the
luminosity equals to the Eddington one $L=L_{\rm Edd}$, and
$L_{\rm Edd}\propto {\frak M}_{\rm H}\propto {\frak M}^h$ (here
${\frak M}_{\rm H}$ stands for the mass of the black hole in
the galaxy center, ${\frak M}$ is the galaxy mass, $\lambda=h$).
The combined case $L=\min(B\Delta m,L_{\rm Edd})$ which also was
considered there corresponds formally to a function $L({\frak
M}_1,{\frak M}_2)$ which is described by two different expressions
of the form (\ref{2.1}), with different $\lambda_{1,2}$, in
different regions. Asymptotics of $I(L)$ for $\lambda_1>0$,
$\lambda_2>0$ can be calculated similarly to how it was done in
Kontorovich, Krivitsky (\cite{kk95pazh}) for $\lambda_1=0$ or
$\lambda_2=0$. Here we give the result, without the derivation,
for completeness:
\begin{equation}
\eqalign{
\gamma=\cases{1-(u+2-2\alpha)/\lambda,&$k<0$,\cr
1-(u+2-2\alpha)/\lambda-k/\lambda_2,&$k>0$,~$L\ll L({\frak M}_0,\mu)$,\cr
1-(u+2-2\alpha)/\lambda+k/\lambda_1,&$k>0$,~$L\gg L({\frak M}_0,\mu)$,}\\
{\rm where} \qquad
k=-\lambda_2(u-2\alpha+2-\lambda)/\lambda-\alpha+u_2-\lambda_2+1.}
\label{sl3}
\end{equation}
For $k>0$, the plot of $I(L)$ has a break; a more flat region
change to a more steep one at $L\sim L({\frak M}_0,\mu)$.

Now we shall consider $I(L)$ for the isotropic distribution
(\ref{isotr}) and with the orbital momentum taken into account.
Unlike (\ref{anis}), for (\ref{isotr}) with ${\bf
J}$ the asymptotics cannot be determined analytically. The reason
is that the dimension of integral (\ref{ilisotr}) for
(\ref{isotr}) is much higher then for (\ref{anis}), due to the
$\delta$-function in~(\ref{anis}). In the same time, there are
some heuristic arguments which lead to the supposition that the
results mentioned above will not essentially change. Indeed, for
(\ref{anis}) $L$ was completely determined by the masses, and the
$\delta$-function in the integral cut a one-dimensional
integration path in the $({\frak M}_1,{\frak M}_2)$ plane (fig.~2
in Kontorovich, Krivitsky, \cite{kk95pazh}). In the case of
(\ref{ilisotr}) $L$ depends not only on masses, but, averaging
over momenta, we can introduce $\overline{L({\frak M}_1,{\frak
M}_2)}$. Due to the scattering of the momenta, not only a
one-dimensional line, but also a whole band close to the line will
make a contribution to the integral over ${\frak M}_1$, ${\frak
M}_2$ (after averaging over ${\bf S}_1$, ${\bf S}_2$, ${\bf J}$).
However, if $\overline{L({\frak M}_1,{\frak M}_2)}$ can be still
described by an expression of the form (\ref{2.1}) then one
may expect that the asymptotics of the integral will not change.

To verify this supposition, we carried out numerical Monte
Carlo simulation. Momentum distribution was taken in the form
(\ref{isotr}), mass distribution $\Phi({\frak M})$ was assumed
to be a Schechter function (\ref{sheh}) for ${\frak M}>{\frak
M}_0$ and $\Phi({\frak M})=0$ for~${\frak M}<{\frak M}_0$. A root
mean square momentum was assumed proportional to the mass in the
power $(3+\beta)/2$ (such a dependence is formed by mergers in the
case $U\propto({\frak M}_1+{\frak M}_2)\*({\frak M}_1^\beta+{\frak
M}_2^\beta)$ and is close to the really observed one, see
discussion in Kontorovich et~al. (\cite{khod}); Krivitsky and
Kontorovich (\cite{kk97})). The luminosity was calculated as
$L=\min(B\Delta m,L_{\rm Edd})$, where $m=S/\sqrt{G{\frak M}R}$.
Unlike the previous section, the orbital angular momentum was
taken into account too.

In this case we may expect that an effective value of $\lambda$
will be~1, $\lambda_1=1$, $\lambda_2=0$ (as we mentioned earlier).
Indeed, for ${\frak M}_1\ll {\frak M}_2$ average proper momenta
are $\overline{{\bf S}_1^2}\propto {\frak M}_1^{3+\beta}$,
$\overline{{\bf S}_2^2}\propto {\frak M}_2^{3+\beta}$, the orbital
one\footnote{The orbital momentum
$\overline{J^2}=\overline{({\frak M}_1{\frak M}_2/({\frak
M}_1+{\frak M}_2))^2v^2p_\infty^2}$, the impact parameter
$\overline{p_\infty^2}\sim(R_1+R_2)^2v_{\rm g}^2/v^2$, so,
$\overline{J^2}\propto {\frak M}_1^2{\frak M}_2^{1+\beta}$.} is
$\overline{{\bf J}^2}\propto ({\frak M}_1v_{\rm
g}(R_1+R_2))^2\propto {\frak M}_1^2{\frak M}_2^{1+\beta}$, after
the merger $\overline{{\bf S}^2}=\overline{{\bf S}_1^2}+
\overline{{\bf S}_2^2}+\overline{{\bf J}^2}\propto {\frak
M}_2^{3+\beta}\left(1+O\left(\frac{\textstyle{\frak
M}_1^2}{\textstyle{\frak M}_2^2}\right)\right)$,
$\overline{m}=\frac{\textstyle\sqrt{\overline{{\bf
S}^2}}}{\textstyle({\frak M}_1+{\frak
M}_2)^{1+\beta}}=m_2\left(1-O\left(\frac{\textstyle
m_1}{\textstyle m_2}\right)\right)$, $\Delta m=m_1+m_2-m\propto
m_1$.

We used the following simulation algorithm (a simplified
description):
\begin{enumerate}
\item\label{i4.1}Two random numbers, ${\frak M}_{1,2}$,
distributed according to the given MF $\Phi({\frak M})$, were
simulated.
\item\label{i4.2}Two random vectors, ${\bf S}_{1,2}$, with
distribution (\ref{isotr}), were simulated.
\item\label{i4.3}Galaxies $1,2$ merged with the probability
proportional to $U({\frak M}_1,{\frak M}_2)$.
\item\label{i4.4}According to the merger cross-section assumed
in our work (see page \pageref{sigma}), the impact parameter and
the relative velocity were simulated, then the merger orbital
momentum ${\bf J}$ was computed.
\item\label{i4.5}The black hole mass ${\frak M}_{\rm H}$ was
simulated (variant~1: ${\frak M}_{\rm H}=\zeta {\frak M}^h$;
variant~2: ${\frak M}_{\rm H}$ is an independent random value with
a power-law distribution).
\item\label{i4.6}Using ${\frak M}_1$, ${\frak M}_2$,
${\frak M}={\frak M}_1+{\frak M}_2$, ${\bf S}_1$, ${\bf S}_2$,
${\bf J}$, ${\bf S}={\bf S}_1+{\bf S}_2+{\bf J}$, ${\frak M}_{\rm
H}$ the luminosity of the active object $L$ was calculated.
\end{enumerate}
Thus, the algorithm gave a random value as an output, which was
distributed according to the same law as the desired luminosity.
Repeating the computations many times, it is possible to find its
distribution function, i.e., $I(L)$.

The procedure of simulation which was actually used was a bit
different from the simplified scheme given above. In
item~\ref{i4.3}, the simulated galaxies merge with some
probability~$p$; with probability $(1-p)$ they are rejected. The
probability $p$ must be proportional to $U({\frak M}_1,{\frak
M}_2)$, e.g., if $U({\frak M}_1,{\frak M}_2)$ is bounded above by
some ${\frak M}_{\rm max}$, we may chose $p=U({\frak M}_1,{\frak
M}_2)/U_{\rm max}$. However, in our case $U({\frak M}_1,{\frak
M}_2)$ is an unbounded function. Moreover, even if we introduce a
limit galaxy mass ${\frak M}_{\rm max}$ and the corresponding
value $U_{\rm max}$ then $p=U({\frak M}_1,{\frak M}_2)/U_{\rm
max}$ will be very small for majority of galaxies and they will be
rejected. This will cause very large computation time. To overcome
this difficulty, we transformed the integrand in (\ref{ilisotr})
as
\begin{equation}
\eqalign{
\int f_1f_2U_{12}
\delta(L-L_{12})F({\bf J})\,d^3J
\,d{\frak M}_1\,d{\frak M}_2\,d^3S_1\,d^3S_2=
\\\qquad
2\int {\frak M}_1^uf_1f_2\frac{U_{12}}
{{\frak M}_1^u+{\frak M}_2^u}\delta(L-L_{12})
F({\bf J})\,d^3J\,d{\frak M}_1
\,d{\frak M}_2\,d^3S_1\,d^3S_2\\\qquad
(f_1\equiv f({\frak M}_1,{\bf S}_1),\ f_2\equiv f({\frak M}_2,{\bf S}_2),\\
\qquad
U_{12}\equiv U({\frak M}_1,{\frak M}_2),\ L_{12}
\equiv L({\frak M}_1,{\frak M}_2,
{\bf S}_1,{\bf S}_2,{\bf J})).
}
\end{equation}
In item~\ref{i4.3} of the algorithm we took $U({\frak M}_1,{\frak
M}_2)/({\frak M}_1^u+{\frak M}_2^u)$ instead of $U({\frak
M}_1,{\frak M}_2)$, and in item~\ref{i4.1} we took ${\frak
M}_1^u\Phi_1$ and $\Phi_2$ instead of $\Phi_1$ and~$\Phi_2$.

\begin{figure}
\centerline{\epsfbox{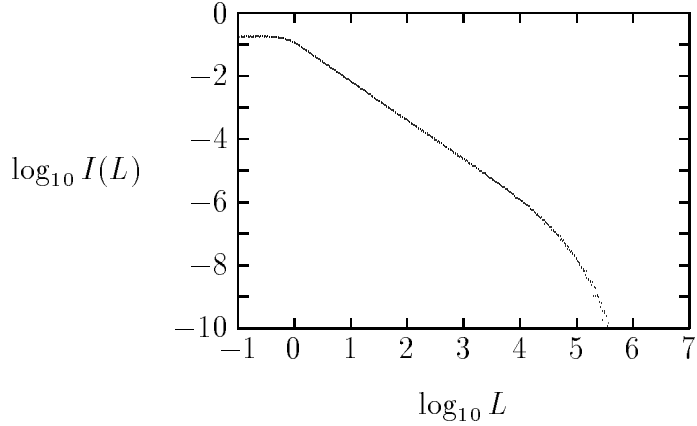}}
\caption[]{An example of the results of numerical simulation for
$I(L)$ (which determine the quasar LF) in the case of the
isotropic momentum distribution and the orbital momentum taken
into account. The luminosity in the figure is given in units $L_1$
(defined in Kontorovich, Krivitsky, \cite{kk95pazh}), $I(L)$ is
normalized to~1. The values of the parameters are: $\mu/{\frak
M}_0=10^6$, $\beta=1/3$, ${\frak M}_{\rm c}/{\frak M}_0=10^3$,
the exponent of the MF is $\alpha=1.25$. The plot confirms the
asymptotic expressions obtained analytically (taking into account
the hypothesis of $\lambda_{\rm eff}=1$, see the text).
As ${\frak M}_0\ll {\frak M}_{\rm c}\ll\mu$,
$u_1+1-\alpha<0$, the formula $I(L)\propto
L^{-1+(u_1+1-\alpha)/\lambda}$ should be applied. For
$\lambda_{\rm eff}=1$ the power-law region of $I(L)$ should be
$L^{-1.25}$, the figure confirms it.}
\label{fig1}
\end{figure}

The results of simulations confirmed the above supposition: the
slope of the power-law region of $I(L)$ coincides with the
predicted value, and $\lambda_{\rm eff}=1=\lambda_1$,
$\lambda_2=0$. As an example, fig.~\ref{fig1} shows the
distribution function for a particular case (see the parameters in
the caption).

Thus, in the most interesting case, when $\lambda_2=0$,
$\lambda_{\rm eff}=1$, $u_1=0$, $u_2=1+\beta$ or 2, the slope of
the power-law region of $I(L)$ (and, therefore, the active objects
LF too) just coincides with the slope of the galaxy MF $\alpha$,
according to equation (16) in Kontorovich, Krivitsky
(\cite{kk95pazh}). This result agrees well with observational data:
according to Binggeli et~al. (\cite{binggeli}); Boyle et~al.
(\cite{boyle}), both $\alpha$ and $\gamma$ are close to~1
(somewhat more).

\section{Luminosity function of active objects}
\label{sec3}
In the previous sections we described rather a general approach
which relates AGN formation to galaxy mergers. Below we shall
consider a concrete application of this approach. The purpose of
this section is to obtain the LF from (\ref{ur1}), (\ref{ur2}) and
find the parameters for which it agrees with the observed one.

We shall take the following input data.

\begin{enumerate}
\item
Galaxy mass function. The bright end of LF (and, so, MF) of normal
galaxies is described well by Schechter's formula (\ref{sheh}).
According to the data obtained in last years, the LF steepens at
its faint end, the slope $\alpha$ reaches $\sim2$ (see, e.g.,
de~Propris et~al., \cite{depropris}; Kashikawa et~al.,
\cite{kashikawa}; Loveday, \cite{loveday}). So, we shall take MF
at $z=0$ as a Schechter function with an additional break:
\begin{equation}
\Phi({\frak M})=\Phi_0(1+({\frak M}/{\frak M}_{\rm
br})^{\alpha_2-\alpha_1}){\frak M}^{-\alpha_2} e^{-{\frak M}/\mu}.
\label{izlom}
\end{equation}
We shall assume $\alpha_1=2$\label{izlom2}, $\alpha_2=1.25$, $\mu$
corresponds to the magnitude $M_B=-21$ (the mass to luminosity
ratio being $\frac {{\frak M}}L\sim10$),
$\Phi_0=5\cdot10^{-3}\mu^{\alpha_2-1}~{\rm Mpc}^{-3}$, and the
break ${\frak M}_{\rm br}$ corresponds to $M_B=-16$.  Possible
change of MF with $z$ will be discussed below.
\item The momentum distribution will be taken in the form
(\ref{isotr}), with the root mean square momentum
\begin{equation}
\frac{\left(\overline{S^2({\frak M})}\right)^{1/2}}
{{\frak M}R\left(\frac{2G{\frak M}}{R}\right)^{1/2}}
={\rm const}=0.1
\label{rmsmom}
\end{equation}
(Krivitsky, Kontorovich, \cite{kk97}).
\item Merger probability. We shall use model (\ref{U}),
assuming $C=\frac{R}{{\frak M}^\beta}=\frac{20{\rm~kpc}}
{(2\cdot10^{11}~{\frak M}_\odot)^\beta}$, $\beta=1/2$, and $v_{\rm
rms}\sim100(1+z)^{-1/2}~{\rm km/s}$, which corresponds to ${\frak
M}_{\rm b}\sim10^{10}(1+z)^{-3/2}~{\frak M}_\odot$ (such a
dependence $v(t)$ takes place in the linear gravitational
instability theory for $\Omega=1$).
\item Active galaxy luminosity. We shall assume $L=\min(L_{\rm
Edd},\varepsilon\eta c^2t_{\rm ac}^{-1}\Delta m)$,
$\varepsilon\sim0.1$; the (bolometric) luminosity is calculated
according to $L/L_\odot=b\cdot10^{0.4(M_{B\odot}-M_B)}$, the
bolometric correction factor being $b\sim10$ (cf. Sanders et~al.,
\cite{sand}; Weedman, \cite{weed}).
\item Black hole mass. Correlation of the masses of black holes
and host galaxies was discussed by Kormendy, Richstone
(\cite{kr})\footnote{Independently, this correlation was
considered in Kontorovich, Krivitsky (\cite{kk95pazh}).} (for
spirals, the bulge masses were taken instead of the galaxy
masses). It was found, that, in average, ${\frak M}_{\rm H}\propto
{\frak M}$.  These results were confirmed by recent HST data
(Press release No.  STScI-PRC97-01). However, there is large
scattering of the ratio ${\frak M}_{\rm H}/{\frak M}$ around its
average value.  Thus, ${\frak M}_{\rm H}\sim2\cdot10^6~{\frak
M}_\odot$ in our Galaxy, whereas ${\frak M}_{\rm
H}\sim3\cdot10^9~{\frak M}_\odot$ in M87.  Below we shall use a
more complicated model than in the previous section: we shall
assume ${\frak M}_{\rm H}=\zeta {\frak M}$, where $\zeta$ is a
random value the decimal logarithm of which is distributed
uniformly in the range $-3\pm1$.
\end{enumerate}

It is well known that the observed quasar LF essentially depends
on the redshift: in the past quasars were much brighter than
now (Boyle et~al., \cite{boyle}). There are two points of view on
this fact in the literature. The simplest interpretation is that
active nuclei are comparatively long-lived objects (with lifetimes
of billions years), but the luminosity of each quasar
decreases in the course of time (``luminosity evolution'').
However, this hypothesis encounters some difficulties. In
particular, if we assume that the luminosity of high-$z$ bright
quasars cannot exceed the Eddington limit then the contemporary
active galaxies, as their ``descendants'', must have very massive
black holes ($\gtrsim10^9~{\frak M}_\odot$), which seems to be
ruled out by observational data (e.g., Schmidt,
\cite{schm2}). The other point of view is that quasars are a
comparatively short ($\lesssim10^8$ years) evolution stage of the
majority of galaxies. Thus, Haehnelt and Rees (\cite{rees}) assume
that an active nuclear grows in almost every galaxy just after its
formation, the initial luminosity equals the Eddington one and so
the LF reflects the MF of black holes. The physical reason of
decreasing the luminosity, proposed by these authors, is that more
massive galaxies form later, their density is lower, the central
potential well is less deep, and the black hole forming there is,
in average, less massive. So, negative correlation between the
mass of the newly formed galaxy and the initial mass of the black
hole in its center is assumed.\footnote{Siemiginowska and Elvis
(\cite{se}) also suppose that active black hole masses
decrease with time.}. Small and Blandford (\cite{bland}) assume
that the observed break in the quasar LF is associated with the
transition between the two modes of accretion: the continuous one
($L=L_{\rm
Edd}$) and the intermittent one ($L<L_{\rm Edd}$ and is controlled
by the amount of ``fuel''). So, decreasing of the break luminosity
is associated with decreasing of the average amount of ``fuel''.
Note that both Haehnelt and Rees (\cite{rees}) and Small and
Blandford (\cite{bland}) assume that the lifetime of an
individual quasar is much less than a characteristic evolution
time.

How can the evolution of the quasar LF be described in the merger
model? In this model the characteristic luminosity corresponding
to the break is related to the mass $\mu$ in (\ref{sheh}),
(\ref{izlom}). Since less massive galaxies form earlier, $\mu$
cannot decrease with time. So, cosmological evolution of
$\Phi({\frak M})$ cannot be the cause of the decreasing of the
quasar luminosity. One of the possible explanation (which we shall
assume in this work) is cosmological evolution of $\eta$
(fraction of mass which actually gets into the center). Here
quasar lifetime is assumed much less than the age of the Universe,
so we may use (\ref{ur1}),
(\ref{ur2}). We shall take $t_{\rm
act}\sim10^8$~years. Cosmological decreasing of $\eta$ may be
caused, for example, by decreasing of the amount of gas in
galaxies.  Indeed, gas and stars behave in different ways at
merging; the matter which gets into the center is, mainly, gas
(Hernquist, Barnes, \cite{gas-to-center}).

To obtain the quasar LF for such a high $z$ as~2, we must take
into consideration cosmological evolution of
$\Phi({\frak M})$. Possible reasons of such evolution are galaxy
mergers, birth of new galaxies of gas, etc. Reliable observational
data on the normal galaxy LF are available only for moderate
redshifts ($z\lesssim0.5$), and the main contribution to the
change in this LF seems to be given by the change of the star
formation rate rather than alteration of the MF (e.g., Small et~al.,
\cite{ssh}). Thus, we have to take the time evolution of
$\Phi({\frak M})$ from model theoretical calculations. The
existing theories for galaxy formation cannot yet give a detailed
and reliable description of this process. However, they make some
qualitative predictions. The evolution of the LF was considered,
e.g., by Kauffmann et~al. (\cite{kauf94}); Cole et~al.
(\cite{cole94}). They concluded, in particular, that an average
galaxy luminosity at $z\sim2$ is several times less than at $z=0$,
whereas the amount of dwarf galaxies is several times higher. In
the same time, both groups of authors notice that they cannot
account for the observational data for both faint galaxy counts
and the slope of the galaxy LF simultaneously: the calculated LF
has much higher $\alpha$ than the observed one. Gnedin
(\cite{gnedin}) obtained the value of $\alpha$ and the shape of
the MF which agree very well with the observed ones (may be,
except for the observed steepening at very small masses) but he
did not compute the time evolution of $\Phi({\frak M})$.

In this paper we shall compute $\phi(L,t)$ for two variants of
$\Phi({\frak M},t)$: 1.~non-evolving $\Phi({\frak M})$ in the form
(\ref{izlom}); 2.~``maximal'' evolution of $\Phi({\frak M})$. In
the second variant we shall take a composed MF and assume
that the shape of $\Phi({\frak M})$ is described by (\ref{izlom}),
$\mu$ and $\Phi_0$ depend on time approximately as in
fig.~3 by Kauffmann et~al. (\cite{kauf94}) and fig.~19 (right
bottom) by Cole et~al. (\cite{cole94}), but, in the same time,
$\alpha_1$ and $\alpha_2$ has the observed values (see page
\pageref{izlom2}), i.e., the MF is more flat than in Kauffmann
et~al. (\cite{kauf94}) and Cole et~al.
(\cite{cole94})\footnote{We have chosen the cases of the fastest
evolution of the MF, and neglect the difference between the
MF and the LF evolution.}.  Namely, we assume
$\mu\propto(1+z)^{-5/3}$, $\Phi_0\propto(1+z)^{4/3}$. The break
mass ${\frak M}_{\rm br}$ is assumed to be constant, as there are
no data for its evolution.

\begin{figure}
\centerline{\epsfbox{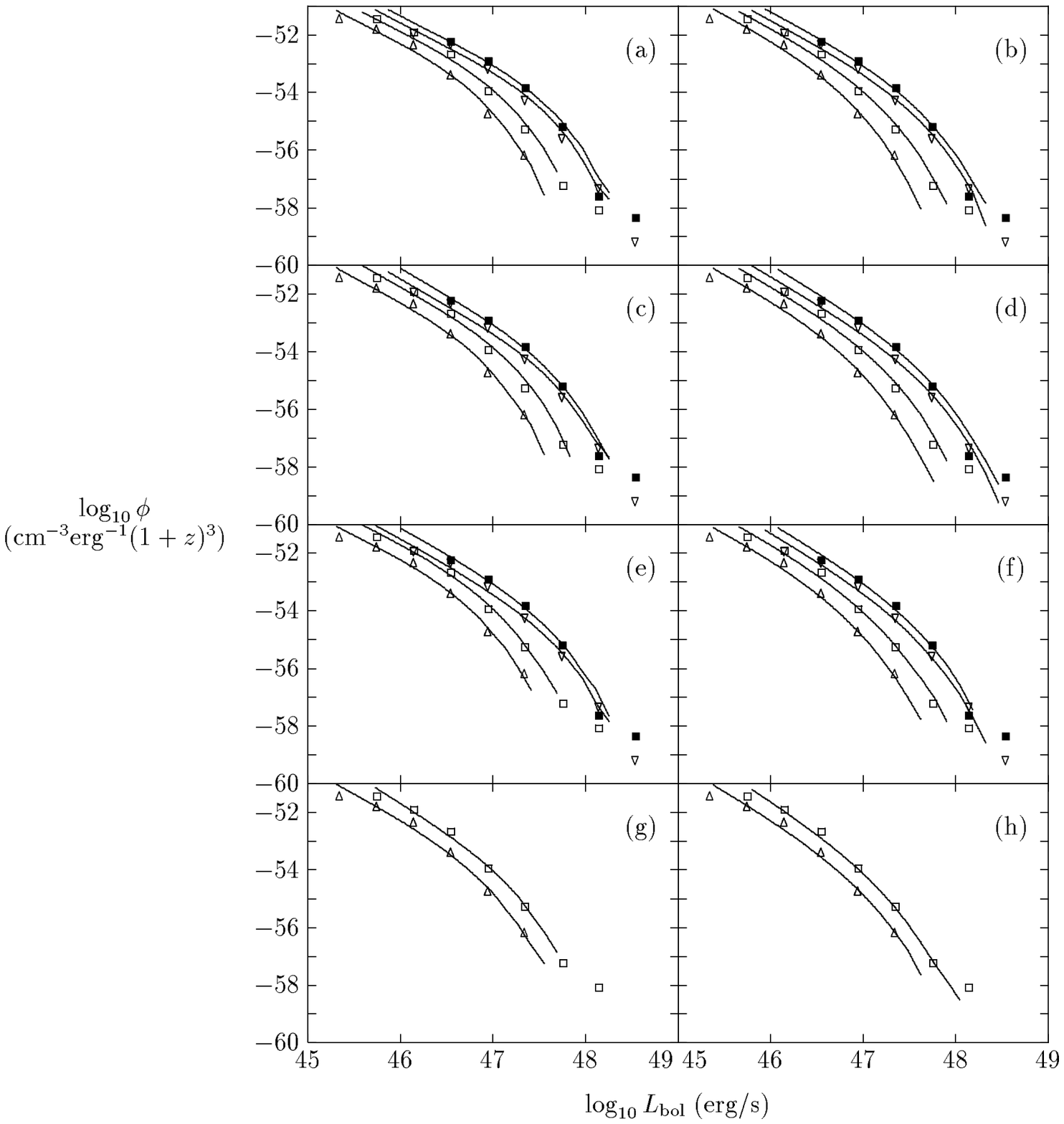}}
\caption[]{Quasar luminosity function: observed (Boyle et~al.,
\cite{boyle}) and predicted by the model described in
section~\ref{sec3}. The corresponding values of the parameters
$\eta$ and $\kappa$ are shown in fig.~\ref{fig3}.
Symbols \dimaplota, \dimaplotb, \dimaplotc, \dimaplotd{} stand for
observational data for $z=0.3$--0.7, 0.7--1.2, 1.2--1.7, 1.7--2.2
(with $q_0=0.5$), solid lines show the merger model results for
$z=0.5$, 0.95, 1.45, 1.95. Figures (a)--(d) show the results for
no Eddington restriction. The left panel corresponds to
(\ref{ur1}), the right one to (\ref{ur2}); Figs. (a) and (b)
correspond to non-evolving mass function,
(c) and (d) are the same for the evolving mass
function. Figures (e)--(h) show the same with the Eddington
restriction. Figures (g) and (h) do not show the plots for large
$z$, see the text.}
\label{fig2}
\end{figure}

\begin{figure}
\centerline{\epsfbox{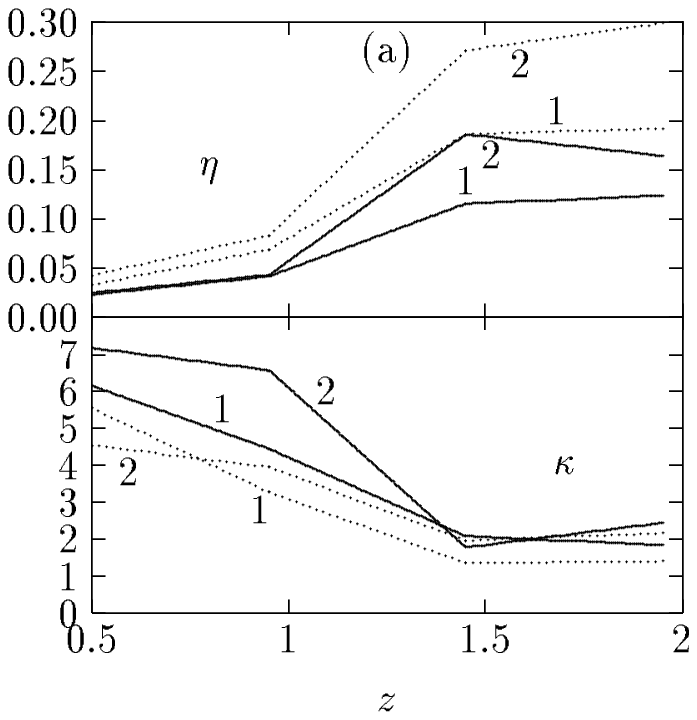}~~~~~\epsfbox{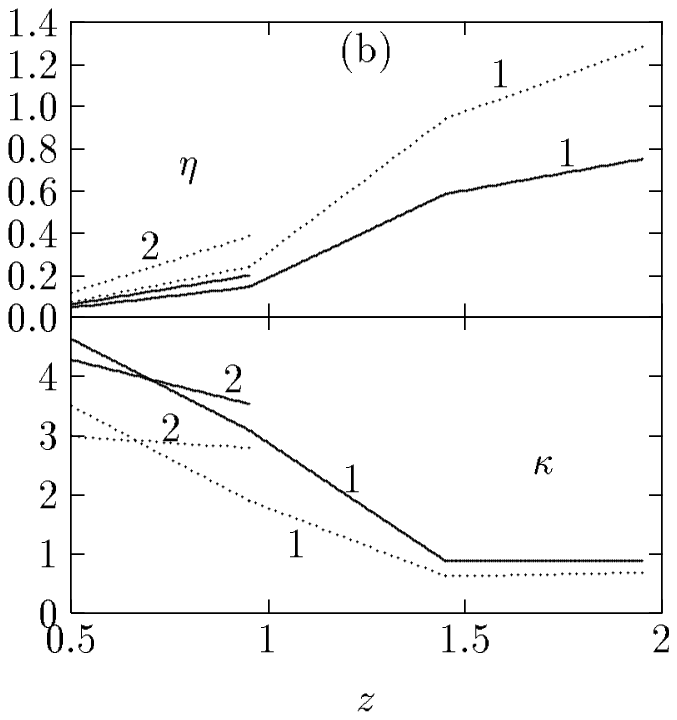}}
\caption[]{Mass fraction $\eta$ which gets into the center and
density contrast $\kappa$, necessary for agreement between
the predicted and observed luminosity functions (fig.~\ref{fig2}).
Solid line corresponds to (\ref{ur1}), dotted line to
(\ref{ur2}), 1 stands for the case without the allowance for the
Eddington restriction, 2 for the case with the restriction, (a) is
for non-evolving mass function, (b) is for the evolved one.}
\label{fig3}
\end{figure}

As merging occurs, mainly, in higher density regions, we should
take into account the non-homogeneity of the galaxy spatial
distribution. Due to this inhomogeneity, $I(L)$ depends on
coordinates, and the right-hand part of (\ref{ilisotr}) will
contain a factor $(\rho/\langle\rho\rangle)^2$, where
$\langle\rho\rangle$ is the average galaxy density (because the MF
will be $\rho/\langle\rho\rangle$ times higher). Integrating
$I(L)$ over a large volume~$V$, we obtain that the average density
of quasars\footnote{Here we average the equation for $\phi$ (or
(\ref{ur1}), (\ref{ur2})) and keep the notation $\phi$ for the
averaged LF.} is
$\displaystyle\kappa=\frac{\int\rho^2\,dV}{\langle\rho\rangle^2V}$
times higher as compared to the homogeneous situation, and $\kappa$
can be expressed as
\begin{equation}
\kappa=\frac{\int\rho^2\,dV}{\langle\rho\rangle^2V}=
\frac{\int\rho\,d{\cal M}}{\langle\rho\rangle{\cal M}}=
\frac{\int\rho\,dN}{\langle\rho\rangle N},
\end{equation}
where $d{\cal M}=\rho\,dV$ is the mass in the volume $dV$, $dN$ is
the number of galaxies in this volume, ${\cal
M}=\langle\rho\rangle V$ and $N$ are the total mass and number of
galaxies. Below the quantity
$\displaystyle\kappa=\frac1N\int\frac\rho{\langle\rho\rangle}\,dN$
will be referred to as the average density contrast.

We will use two $z$-dependent fitting parameters in comparing the
quasar LF with the observed one: the fraction of matter
$\eta$ and the average density contrast~$\kappa$.

The results are shown in figs.~\ref{fig2} (the AGN LF) and
\ref{fig3} (the corresponding values of the parameters for which
$\phi(L)$ has best agreement with the observational data by Boyle
et~al. (\cite{boyle}) for the quasar LF). The figures show that the
model presented above is able to account for the observed
evolution of $\phi(L)$. In the case of non-evolving MF the
fraction of matter which gets into the center $\eta$ changes from
$\approx0.12\mbox{--}0.3$ for $z\approx2$ to
$\approx0.025\mbox{--}0.043$ for $z\approx0.5$, whereas the
average density contrast in the regions of quasar formation
$\kappa$ is
$\approx1.4\mbox{--}2.4$ for $z\approx2$ and
$\approx4.5\mbox{--}7.2$ for $z\approx0.5$.
In the case of ``maximal'' MF
evolution the parameter $\eta$ for large redshifts is much higher:
$\approx0.8\mbox{--}1.3$ for $z\approx2$ (see the discussion
below), $\approx0.05\mbox{--}0.11$ for
$z\approx0.5$, and $\kappa$ is somewhat lower:
$\approx0.7\mbox{--}0.9$ for $z\approx2$, $\approx3\mbox{--}4.6$
for $z\approx0.5$. Taking into account the Eddington restriction
gives in most cases an increase of $\eta$ and $\kappa$;
for (\ref{ur2}) $\eta$ is somewhat higher, whereas
$\kappa$ is somewhat lower, as compared to (\ref{ur1}).

Alteration of $t_{\rm ac}$, $\varepsilon$, $b$ results in
alteration of $\eta$ and $\kappa$, according to
\begin{equation}
\eta\propto t_{\rm ac}\varepsilon^{-1}b,\qquad
\kappa\propto t_{\rm ac}^{-1}.
\label{scaling}
\end{equation}
Besides, $\eta$ increases with decreasing the average angular
momentum in (\ref{rmsmom}).

In some cases (namely, the evolving MF and (\ref{ur2})) the values
given in fig.~\ref{fig3} fall beyond the physically allowed range
(the fraction of matter which falls into the center cannot
exceed~1, and the density contrast must be higher than~1). It does
not mean that the model fails: according to (\ref{scaling}), the
values falls into the required range if, for example, $t_{\rm ac}$
is $5\cdot10^7$~years instead of $10^8$.

Note that the obtained $\eta$ values for $z\approx0.5$ (several
per cent) have the same order of magnitude as an average gas
fraction in modern galaxies, $\kappa\approx10$
corresponds to an average density contrast in the large-scale
structure filaments, and the value $z\sim1$ corresponds to the
epoch of intensive star formation accompanying by decreasing of
the amount of gas. Next, if the parameter $\eta$ is really related
to the gas fraction in galaxies, then such high $\eta$ values for
$z\sim2$ as several tenth looks rather natural: a large quantity
of gas has not yet turned into stars. However, for galaxies with
such a high gas fraction the merger criterion described in
section~\ref{discus} should be modified, as well as the expression
for $U$, because in a high-speed collision ($v\gg v_{\rm g}$) two
stellar systems will pass through each other,
whereas two colliding gas clouds will show quite a different
behavior, forming a dissipative discontinuity system (e.g.,
Chernin, \cite{cher}).

In the variant with the ``maximal'' MF evolution and the Eddington
luminosity taken into account, the obtained LF at large redshifts
($z\gtrsim1.5$) disagrees with the data by Boyle et~al.
(\cite{boyle}) for any values of the parameters: the number of the
brightest quasars ($L\sim10^{48}~{\rm erg/s}$) is much lower than
the observed one (that is why the last two curves are not shown in
figs.~\ref{fig2}g and \ref{fig2}h). It is related to the influence
of the Eddington restriction. Indeed, if we assume $L\le L_{\rm
Edd}$, then such quasars must have a black hole of a mass ${\frak
M}\sim10^{10}~{\frak M}_\odot$. In the same time, the ``maximal''
MF evolution assumed here implies that the masses of galaxies at
$z\sim2$ are approximately one order lower than the modern ones,
whereas the black hole masses ${\frak M}_{\rm H}=\zeta{\frak M}$.
Thus, there are too few black holes with ${\frak
M}\sim10^{10}~{\frak M}_\odot$, even in spite of the $\zeta$
scattering assumed here. There are many possible reasons for this
discrepancy: 1.~the luminosity may be much higher than the
Eddington one due to anisotropy of the quasar ``central machine'';
2.~in a model where periods of activity alternate with pauses, the
peak luminosity may be higher than the Eddington one;
3.~gravitational lensing may cause an increase of the apparent
brightness; 4.~the case for ``maximal'' MF evolution may not be
realized. Allowing for the former two factors should result in an
increase of the luminosity (therefore, lower $\eta$ required) and
decrease of the normalization (higher $\kappa$
required), which gives one more explanation of the curves in
fig.~\ref{fig3}b which falls beyond the allowed region.

Thus, the merger model can explain the observed shape and
evolution of the quasar LF and give an estimate for the parameters
$\eta$ and $\kappa$. Also, in the case of evolving MF
the results agree well with the hypothesis that the quasar
luminosity may much exceed the Eddington limit.

\end{document}